\documentclass[%
preprint,
showpacs,preprintnumbers,
 amsmath,amssymb,
 aps,
prb,
]{revtex4-1}

\usepackage{graphicx}
\usepackage{dcolumn}
\usepackage{bm}

\begin{document}


\title{Role of fluctuations and non-linearities on field emission nanomechanical self-oscillators}

\author{T. Barois}
\author{S. Perisanu}
\author{P. Vincent}
\author{S. T. Purcell}
\author{A. Ayari}%
 \email{anthony.ayari@univ-lyon1.fr}
\affiliation{%
Institut Lumi\`ere Mati\`ere, UMR5306 Universit\'e Lyon 1-CNRS, Universit\'e de Lyon 69622 Villeurbanne cedex, France.
}%


%

\date{\today}

\begin{abstract}
A theoretical and experimental description of the threshold, amplitude and stability of a self-oscillating nanowire in a field emission configuration is presented. Two thresholds for the onset of self-oscillation are identified, one induced by fluctuations of the electromagnetic environment and a second revealed by these fluctuations by measuring the probability density function of the current. The AC and DC components of the current and the phase stability are quantified. An AC to DC ratio above 100\% and an Allan deviation of 1.3$\cdot$10$^{-5}$ at room temperature can be attained. Finally it is shown that a simple non-linear model cannot describe the equilibrium effective potential in the self-oscillating regime due to the high amplitude of oscillations.
\end{abstract}
\maketitle


\section{\label{sec:level1}INTRODUCTION}

Research on Nanoelectromechanical systems (NEMS) has recently reached several important milestones in sensing \cite{chaste2012} and quantum physics \cite{oconnell2010}. In addition, the non linear properties \cite{cross2008,PhysRevB.81.165440,nanoletnonlin2008,VDZPRL2010} as well as the comprehension of the dissipation mechanisms of NEMS \cite{PhysRevLett.105.027205,eichler_nonlinear_damping} is attracting increasing interest. Although negative intrinsic damping in NEMS, i.e. self-oscillation\cite{Jenkins2012} with nanoscale feedback has already been observed \cite{1364061,ayari_self_oscillations,grogg,kleshch2010,Steeneken2011} an experimental study of its non-linear nature is still missing. The non-linear terms are crucial for a stable self-oscillator because they govern the amplitude of the AC output. In fact, these terms must contain a coefficient with the appropriate sign in order to reach a saturation regime where a stable limit cycle can form, otherwise the system amplitude might diverge. Moreover, depending on the sign of these coefficients a self-oscillating system can be either supercritical where it is possible to pass continuously from an immobile behavior to a self-oscillation regime, or subcritical with an abrupt jump to a self-oscillating state and hysteresis. A sub-critical self-oscillator is usually more non linear meaning a less pure output signal and more harmonics. A supercritical self-oscillator can be tuned in amplitude output down to zero while a subcritical one cannot.

Sub-critical self-oscilllation of a field emission NEMS was first observed in Ref.~\onlinecite{ayari_self_oscillations} in a bottom-up geometry with nanowire resonators (NWRs). In our previous investigations of the self-oscilllation of field emission NEMS, we focused on the description of a theoretical criterion to predict the linear instability \cite{lazarus_simple2010} and a more detailed numerical analysis of the non-linear behavior in a model geometry was performed.\cite{lazarus_statics2010} In this article, we will compare new and extensive experimental results with the linear and non-linear predictions for two SiC NWRs. The experimental set-up and the direct evidence of self-oscillation are shown in section \ref{lin}. Section \ref{undirect} presents in detail the theoretical criterion that determines the self-oscillation threshold of our system, as well as a simplified and less obscure model. Then the method to test this model is presented which appears to fail to predict the threshold measured in section \ref{lin}. This failure comes from two reasons : i) experimental uncertainties to determine accurately all the physical parameters; ii) the existence of 2 thresholds in the system. This last point is explained and experimentally confirmed by studying AC current fluctuations in section \ref{nonlin}. Finally, we propose a basic theoretical description of the non-linear dynamics of our system and use this to analyze the amplitude and phase stability of our self-oscillator.
\begin{figure}
\includegraphics[width=7cm]{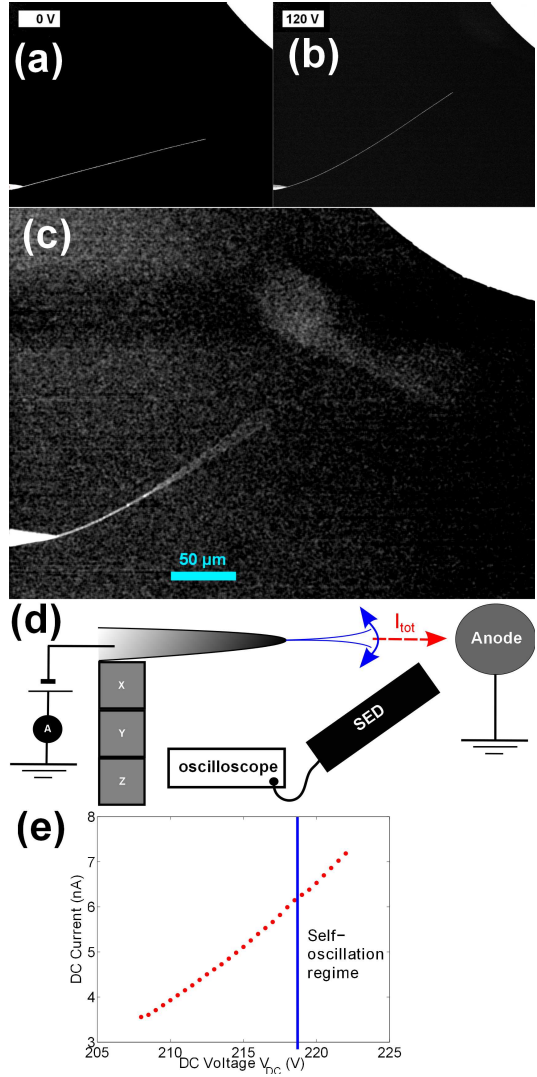}
\caption{\label{fig:sample1}(a) SEM image of sample 1 for 0 V. (b) SEM image of sample 1 bent by electrostatic forces. (c) SEM image of sample 1 in self-oscillation. (d) Schematic diagram of the experimental set-up. (e) Field emission I$_{DC}$-V$_{DC}$ curve for sample 1 with the SEM beam OFF and a different configuration compared to (a),(b) or (c) (i.e. with a higher threshold voltage) . The vertical line separates the region with self oscillation from the region without self-oscillation.
}
\end{figure}

\section{DIRECT EXPERIMENTAL DETERMINATION OF THE THRESHOLD}
\label{lin}
The experiment takes place in an ultra high vacuum chamber with a scanning electron microscope (SEM). It consists of a NWR attached to a tungsten tip positioned with an XYZ piezoelectric motor, in front of a metallic anode connected to the ground (see Table \ref{tabrecap} for samples size and figure \ref{fig:sample1}(a) for a SEM image of sample 1). The tip is at a negative voltage with respect to the ground. A Keithley 6517 electrometer provides the DC voltage and records the DC current due to field emission at the apex of the NWR. The emitted electrons from the apex are attracted by the anode and generate secondary electrons collected by the secondary electron detector (SED) of the SEM chamber as shown on figure \ref{fig:sample1}(d). The voltage output of the SED is recorded on a 1 GHz bandwidth oscilloscope (4 millions data points acquisition) so that the measurement is only limited by the SED bandwidth ($\sim$ MHz) and is proportional to the total current DC plus AC. The voltage SED is then calibrated and converted into current thanks to the average DC current measured by the electrometer for different value of the DC voltage. The NWR is manoeuvered in the vicinity of the anode with a piezoelectric motor, to find a favorable configuration for self-oscillations. In general it requires the NWR to be rather close to the counter electrode (less than 10 $\mu$m) and bent by electrostatic forces as shown in figure \ref{fig:sample1}(b), although we observed self-oscillations sometimes in an apparently symmetric position. Spontaneous oscillations in the transverse direction are observed by SEM imaging when the DC voltage is above a threshold voltage (see figure \ref{fig:sample1}(c)).

After the first determination of the self-oscillation conditions, I$_{DC}$-V$_{DC}$ curves were measured with the SEM beam OFF. In Fig.~\ref{fig:sample1}(e), one notices that in contrast to what we reported in Ref.~\onlinecite{ayari_self_oscillations}, the field emission DC current can reach a self-oscillating regime without DC current jumps and hysteresis. It is tempting to believe that a geometry with a supercritical transition has been obtained. However in the following it will be shown that the transition is still sub-critical (i.e with a discontinuity in vibration amplitude as a function of applied voltage). In fact, the jump in DC current becomes so small that it is below the noise level. However another more important proof of a discontinuous response is still measurable and will be presented below.

\begin{figure}
\includegraphics[width=7cm]{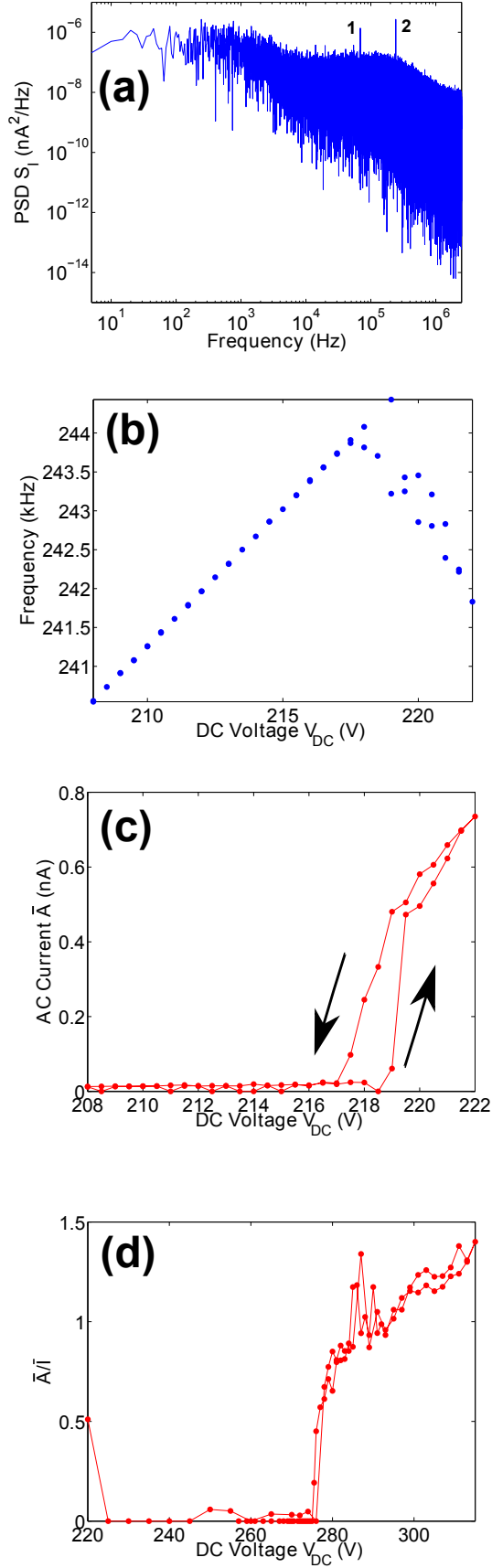}
\caption{\label{fig:A(Vdc)} (a) PSD of the total current for sample 1 at V$_{DC}$ = 213 V. The peak labeled 1 (respectively 2) is the first (respectively second) mode of the NWR. (b) Self-oscillation frequency versus DC voltage from the PSD of the total current for sample 1. (c) Amplitude $\bar{A}$ of the AC current versus DC applied voltage for sample 1. The arrows indicate the voltage sweep direction. (b) Ratio of the AC current to the DC current (i.e. $\bar{A}/\bar{I}$)  versus voltage for sample 2.}
\end{figure}

For each DC voltage, the SED signal is typically recorded for 0.2 s in order to have about 100 points per period and determine the self-oscillation AC current amplitude with accuracy. In our experiment, the time dependent field emission total current is measured rather than the position x(t) to study the NWR. The field emission total current is a complex transduction of the motion of the NWR as it depends non-linearly on the apex voltage as well as the NWR position. The dependence of the current from the position comes from the field enhancement factor. This term depends on the electrostatic geometry such as the NWR radius, the distance between the NWR apex and counter electrode or the tungsten tip. This geometry is either fixed or determined by x(t), so it can be described by a single parameter the position x(t). Below the self-oscillation threshold, the thermomechanical noise of the  NWR cannot be detected in the SED signal because the thermal noise is too small (see section \ref{verifexp}). However the electrical noise has a white component high enough to reveal the resonance in the power spectrum density (PSD) of the total current (see fig. \ref{fig:A(Vdc)}(a)). Figure \ref{fig:A(Vdc)}(b)  represents the variation of this resonant frequency versus DC voltage corresponding to the $\delta \omega$ term in the equation \ref{Eqsimpl} that will be introduced in the next section. It shows that it increases linearly with the voltage until a sudden slope sign change. This linear dependence is expected for instance from electrostatic tuning \cite{PhysRevB.77.165434,PhysRevLett.89.276103} and in such a narrow voltage range. The voltage, where this slope change occurs, corresponds to the beginning of self-oscillation. Measurements are performed with 0.5 V steps. For sample 1, the frequency increase linearly up to 218.5 V and at 219 V it deviates significantly from this trend. So the self-oscillation threshold is at 218.75 $\pm 0.25$ V and similarly at 276.25 $\pm 0.25$ V for sample 2.

In the self-oscillation state, the time dependent current, in its simplest form, is given by $I_{tot}(t)-\bar{I} = I(t) = A \cos(\omega t -\varphi)$ where $I_{tot}(t)$ is the total field emission current, $\bar{I}$ is the DC current, A is the self-oscillation (i.e. AC) current amplitude, $\omega$ the self-oscillator angular frequency and $\varphi$ its phase. A(t) and $\varphi(t)$ are the two slowing varying degrees of freedom compared to the period of the self-oscillator. Their dynamics are described by two different differential equations. A(t) and $\varphi(t)$ can be experimentally obtained from the filtered total current signal with the help of a Hilbert transform\cite{pikovsky2003synchronization} :
\begin{eqnarray}
I_H(t) = I(t)*\frac{1}{t} = p.v.\int_{-\infty}^{+\infty}\frac{I(\tau)}{t-\tau}d\tau\\
Ae^{i\varphi} = I(t) + i I_H(t)
\end{eqnarray}
where $p.v.$ is the principal value. Figure \ref{fig:A(Vdc)}(c) gives $\bar{A}$ the average value of A for different voltages. The sudden increase of $\bar{A}$ gives the same self-oscillation threshold as the one from the resonant frequency.  This abrupt change at 219 V for sample 1 indicates that the transition is subcritical. This can be noticed as well, by sweeping down the voltage while the self-oscillation persists down to 217 V. The hysteresis is clearly seen here while unobservable in the I-V curve. We didn't succeed to decrease the hysteresis below 1V and in some cases this bistability region (i.e. where the self-oscillation solution and the non-oscillating solution coexist) can be higher than 10 V. The I$_{AC}$/I$_{DC}$ ratio increases with the voltage and as observed in Fig. \ref{fig:A(Vdc)}(d), the AC component can be bigger than the DC one.

\section{INDIRECT DETERMINATION OF THE THRESHOLD}
\label{undirect}
The threshold for self-oscillation can be predicted if the dynamical equation describing the system is found and the physical parameters involved are measured. This analysis can identify the most important elements in favor of self-oscillation. The general dynamical mechanical equation was given in Ref.~\onlinecite{ayari_self_oscillations}. Its linearized version around the equilibrium position x$_{eq}$ and $\bar{U}$, the DC voltage difference between the apex and the anode (by convention this voltage is taken positive),  is :
\begin{equation}
\label{Eq1} \ddot{x}+\frac{\omega_0}{Q}\dot{x}+\omega_0^2x= \frac{C'}{m^*}
\bar{U}U
\end{equation}
where x is the apex displacement in the direction perpendicular to the NW (x positive when the NW approach the anode, the sign convention is important here to determine the stability), $C'$ is the spatial derivative in the x direction of the capacitance C, m* the effective mass of the NW, U the AC voltage, Q the quality factor and $\omega_0/2\pi$ the resonance frequency.\\
The mechanical equation is coupled to a linearized electrical equation, obtained from Kirchhoff's law :
\begin{equation}
\label{elec01}
(\frac{\partial I}{\partial
U}+\frac{1}{R_{NW}})U+C\dot{U}=-\frac{\partial I}{\partial
x}x-C'\bar{U}\dot{x}
\end{equation}
where I is the field emission AC current and $R_{NW}$ the nanowire resistance.
\subsection{The Routh-Hurwitz criterion}
Inserting Eq. \ref{elec01} in \ref{Eq1}, we get :
\begin{eqnarray}
C\dddot{x}+\ddot{x}\left(\frac{\partial I}{\partial
U}+\frac{1}{R_{NW}}+C\frac{\omega_0}{Q}\right)+\dot{x}\left[\frac{C'^2\bar{U}^2}{m^*}+\left(\frac{\partial I}{\partial
U}+\frac{1}{R_{NW}}\right)\frac{\omega_0}{Q}+C\omega_0^2\right]\nonumber\\
+x\left[\frac{C'\bar{U}}{m^*}\frac{\partial I}{\partial x}+\left(\frac{\partial I}{\partial
U}+\frac{1}{R_{NW}}\right)\omega_0^2\right] = 0
\end{eqnarray}
The stability of this dynamical system can be checked with the Routh-Hurwitz criterion (see Ref.~\onlinecite{ROK} p. 219), which says that a differential equation of the form $a\dddot{x}+b\ddot{x}+c\dot{x}+dx =0$ has only negative eigenvalues real part if and only if $a >0, b>0, d >0$ and $bc-ad > 0$. So for our system, self-oscillation begins when the following Routh-Hurwitz expression (RHE) becomes positive :
\begin{eqnarray}
\label{EqRH}
C\left[\frac{C'\bar{U}}{m^*}\frac{\partial I}{\partial x}+\left(\frac{\partial I}{\partial
U}+\frac{1}{R_{NW}}\right)\omega_0^2\right]-\left(\frac{\partial I}{\partial
U}+\frac{1}{R_{NW}}+\frac{C\omega_0}{Q}\right)\nonumber\\
\left[\frac{C'^2\bar{U}^2}{m^*}+\left(\frac{\partial I}{\partial U}+\frac{1}{R_{NW}}\right)\frac{\omega_0}{Q}+C\omega_0^2\right] \geq 0
\end{eqnarray}
To fulfill this criterion we showed in Ref.~\onlinecite{lazarus_simple2010} and \onlinecite{lazarus_statics2010} that it is easier but not absolutely necessary to firstly have :
\begin{equation}
\label{EqRj}
\frac{\partial I}{\partial U}\sim\frac{1}{R_{NW}}
\end{equation}
and secondly
\begin{equation}
\label{Eqtps}
\omega_0C/(\frac{\partial I}{\partial U}+\frac{1}{R_{NW}}) = \omega_0RC \sim 1
\end{equation}
where R is the equivalent resistance of the 2 parallel resistances of the circuit, i.e. the field emission resistance and the nanowire resistance.

\subsection{Simplified model}
A drawback of the Routh-Hurwitz criterion is that it obscurs the physical origin of the self-oscillation regime. A less rigorous criterion can be obtained by simply looking for a stationary solution $x(t) = X \cos(\omega_0 t)$ where $X$ is the amplitude of self-oscillation. By inserting this solution into the electrical equation \ref{elec01}, the voltage can be expressed as function of $x$ and $\dot{x}=-X\omega_0 \sin(\omega_0 t)$ :
\begin{equation}
U = \frac{-R\frac{\partial I}{\partial x}-C'R^2C\omega_0^2\bar{U}}{1+(RC\omega_0)^2}x+\frac{-RC'\bar{U}+R^2C\frac{\partial I}{\partial x}}{1+(RC\omega_0)^2}\dot{x}
\end{equation}
Then this expression can be used to replace U in Eq. \ref{Eq1} to obtain:
\begin{eqnarray}
\label{Eqsimpl}
\ddot{x}+\gamma\dot{x}+[\omega_0^2+\delta \omega^2]x= 0\\
\gamma = \frac{\omega_0}{Q}-\bar{U}\Gamma(RC\frac{\partial I}{\partial x}-C'\bar{U})\nonumber
\end{eqnarray}
where $\delta \omega^2$ is the frequency tuning due to the electromechanical coupling $\gamma$ is the effective damping and $\Gamma = \frac{RC'}{m^*[1+(RC\omega_0)^2]}$. Self-oscillation will take place if the damping goes to zero. It requires first, that the term with the spatial derivative of the AC current is higher than the term with $C'$ (the so called electrostatic damping \cite{TB2012}). The first term increases exponentially with voltage, as long as the field emission resistance is not too small compared to the $R_{NW}$, while the term with $C'$ increase roughly linearly. So, if the NWR supports the necessary DC current, the first term can dominate. If this condition is fulfilled then by increasing the DC voltage, self oscillation should occur above a certain threshold. 
As well, close to the threshold the damping should change linearly with voltage. With these notations the RHE can be rewritten as:
\begin{equation}
\frac{\omega_0}{Q}\left(1+\frac{RC\omega_0}{Q}\frac{1}{1+(RC\omega_0)^2}\right)-\bar{U}\Gamma\left(RC\frac{\partial I}{\partial x}-C'\bar{U}(1+\frac{RC\omega_0}{Q})\right) \leq 0
\end{equation}
So the RHE essentially differs from the simplified model by 2 terms that are negligible as long as the relation \ref{Eqtps} is verified and Q is high.
\subsection{Comparison between the model and the experiment}
\label{verifexp}
In principle, the Routh-Hurwitz criterion should give the value of the threshold for self-oscillation. 
However, this voltage is very sensitive to the values of some parameters (see below) and measuring the corresponding experimental parameters with a high accuracy is not always possible due to their DC voltage dependence and the instability of the field emission DC current. It turns out that it is illusive to try to make an accurate prediction of the threshold due to these experimental uncertainties. More importantly, as it will be explained in the next section, more accurate measurements of the experimental parameters couldn't even predict the threshold measured in section \ref{lin}. Nevertheless, we succeeded in obtaining reasonable experimental estimates of all the physical parameters : compared to our first studies\cite{ayari_self_oscillations}, this time mostly all values are measured and not simply guessed from somewhat questionable theoretical considerations. Moreover the model has been qualitatively confirm by varying some experimental parameters and comparing the expected and measured variation in the threshold. Some aspects of the model will also be tested in the next section.\\
We need to measure $ m*, \omega_0, Q, R_{NW}, \frac{\partial I}{\partial
U}, C, C', \bar{U}, \frac{\partial I}{\partial x}$. 

\begin{table}
\caption{\label{tabrecap}%
Physical parameters extracted from various experiments. }
\begin{ruledtabular}
\begin{tabular}{ccc}
  Parameters & sample 1 & sample 2 \\
  \hline
  Length & 220 $\mu$m & 198 $\mu$m \\
  \hline
  Radius & 115 nm &  160 nm \\
  \hline
  Effective mass & 6.9$\cdot$10$^{-15}$ kg & 1.2$\cdot$10$^{-14}$ kg \\
  \hline
  $\omega_0/2\pi$ & $\sim$ 0.25 MHz & $\sim$ 35 kHz \\
  \hline
  Q & 11 000 & 6 000 \\
  \hline
  C' & 0.15 pF/m & 0.47 pF/m \\
  \hline
  $R_{NW}$ & 2.5 G$\Omega$ & 1.5 G$\Omega$ \\
  \hline
  $(\frac{\partial I}{\partial U})^{-1}$ & 1.1 G$\Omega$ & 187 G$\Omega$ \\
  \hline
  C & 4 fF & 1.3 fF\\
  \hline
  $\omega_0$RC & 15 & .42\\
  \hline
  $I_{DC}$ at the experimental threshold & 6.2 nA & 139 pA\\
  \hline
  $\frac{\partial I}{\partial x}$ at the experimental threshold & 310 pA/$\mu$m &  83 pA/$\mu$m\\
  \hline
  Calculated Routh-Hurwitz threshold &  222.6 $\pm$ 0.19 V & 278.2 $\pm$ 0.36 V\\
  \hline
  Calculated simplified threshold &   222.4 $\pm$ 0.16 V & 278.17 $\pm$ 0.36 V\\
  \hline
   Experimental threshold &  218.75  $\pm$ 0.25 V & 276.25 $\pm$ 0.25 V\\
  \hline
  Experimental PDF threshold &  223.3 $\pm$ 0.77 V &  278.4 $\pm$ 0.82 V\\
\end{tabular}
\end{ruledtabular}
\end{table}

Table \ref{tabrecap} sums up the experimental parameters used to calculate the Routh-Hurwitz criterion for the two samples. We estimate the effective mass from the dimensions of the samples from SEM images (see Fig.~\ref{fig:sample1} for sample 1) the SiC density ($3200~kg/m^3$) and first mode correction coefficient of 0.25\cite{PhysRevB.81.165440,APLsorin2011}. The resonant frequency obtained from figure \ref{fig:A(Vdc)}(b) near the self-oscillation experimental threshold gives $\omega_0$. Q is obtained from the measurement in spot mode in a SEM from the mechanical resonance peak width. This measurement is performed at low voltage in order to reduce the electrostatic damping (see Ref.~\onlinecite{TB2012} for details). The resonant frequency of the first mode for different voltages gives the voltage dependent effective rigidity. The resonant frequency at zero voltage is in reasonable agreement with the geometry and the expected Young modulus. Next, the voltage dependence of C' is deduced from the rigidity and the SEM imaging of the nanowire bending with the voltage. Finally, we deduce the resistance $R_{NW}$ from the increase of the resonance width at high voltage measured not too close to the self-oscillation experimental threshold. I$_{DC}$V$_{DC}$ data are then replotted versus $\bar{U} = V_{DC} + R_{NW}\bar{I}$ to get $\bar{I}(\bar{U})$ and $\frac{\partial I}{\partial U}$ after numerical derivation. In the range of voltage of our experiments $\frac{\partial I}{\partial U}$ appears to be rather constant. The capacitance C is obtained by applying a voltage step at the counter electrode and by measuring the characteristic time RC of the field emission total current transient with the SED. \\

$\partial I/\partial x$ is the most important parameter in the damping canceling mechanism and self-oscillation. A first approach to measure this parameter is by oscillating the piezo actuator with a low frequency AC voltage during field emission and measuring with a lock-in the SED signal. The motion of the piezo actuator is then calibrated with the SEM. However, this measurement probably underestimates the actual value of this parameter as this procedure doesn't reproduce properly the direction of motion nor the amplitude. It gives a $\partial I/\partial x$ of several pA/$\mu$m at 1 nA, with such a value self-oscillation are impossible according to the model. A significantly higher value of $\partial I/\partial x$ during oscillation is necessary to get a reasonable agreement with the experiment. A better estimates of $\partial I/\partial x$ consists in imaging the self-oscillation amplitude with the SEM as well as measuring the AC current. $\partial I/\partial x$ is obtained by dividing the amplitude of oscillation by the AC current. This gives a proper order of magnitude to calculate a threshold coherent with what will be measured in the next section. However, due to some uncertainties in the image analysis, the amplitude can not be estimated with an accuracy better than 20 $\%$ which corresponds to a change in the threshold value by several volts, so this method cannot be considered as reliable to precisely predict the threshold. Moreover this measurement can be performed only in a narrow range of voltage: for a voltage above the onset of self-oscillation but for a voltage low enough so that the field emission secondary electrons current is lower than the SEM secondary electrons current (otherwise the SEM image is saturated and appears white). In between, the field emission current from the nanowire and the electron beam current are comparable and both contribute to the signal in the SED. The presence of this additional current from field emission is responsible for the noise in the image and the deterioration of the image analysis accuracy. Theoretically $\partial I/\partial x$ increases quasi linearly with the DC current due to the exponential dependence of the Fowler Nordheim DC current, so we measured it only for a fixed DC current and then included this dependence in the formula.

From this measurement, the amplitude of the expected thermo-mechanical noise at the resonance mentioned in section \ref{lin} can be calculated by :

\begin{equation}
\sqrt{S_{Ixth}}(f_0)=\sqrt{\frac{4k_bTQ}{m\omega_0^3}}\frac{\partial I_{FN}}{\partial
x}
\end{equation}
For the experimental condition of Fig. \ref{fig:A(Vdc)} (a), $\sqrt{S_{Ixth}}(f_0)$ = 18 fA/$\sqrt{Hz}$ whereas the experimental peak is at 1.6 pA/$\sqrt{Hz}$. So the origin of the resonant peak is not the thermomechanical noise. As the nanowire is actuated electrostatically a voltage white noise can induce a peak in the PSD at the resonant frequency. Theoretically, for our sample, the voltage noise due to the field emission shot noise dominates the Johnson noise from the nanowire resistance and gives a peak of amplitude :
\begin{equation}
\sqrt{S_{Ixshot}}(f_0)=\sqrt{2e\bar{I}}R_{NW}\frac{C'U}{2}\frac{Q}{m\omega_0^2}\frac{\partial I_{FN}}{\partial
x} = 211 fA/\sqrt{Hz}
\end{equation}
This value is still an order of magnitude lower than the experimental peak indicating that another source of white noise, that we couldn't identify is responsible for this peak. Observing the thermomechanical noise would require a better quality factor or lower nanowire resistance.
\begin{figure}
\includegraphics[width=8cm]{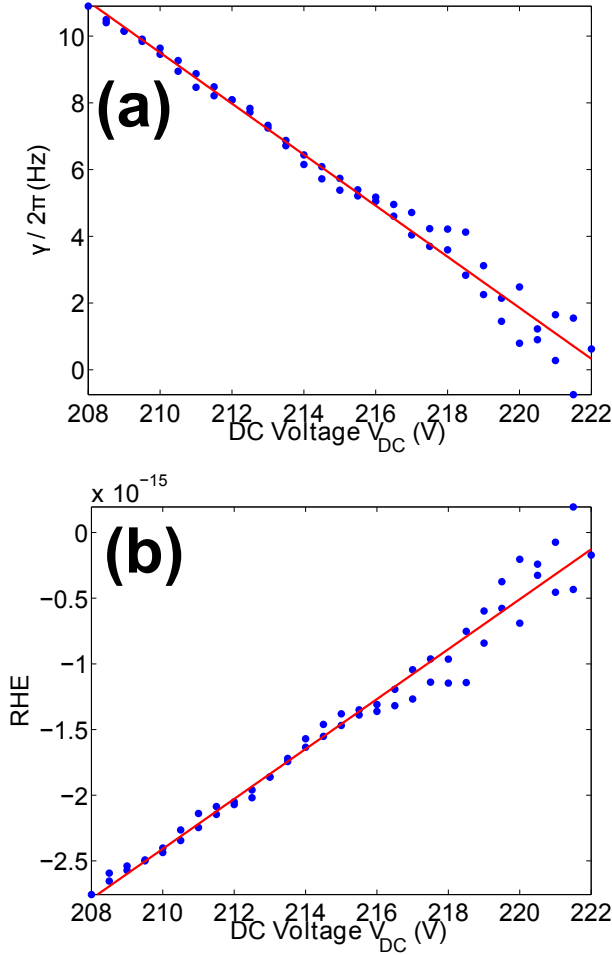}
\caption{\label{fig:RH} (a) Calculated damping from Eq. \ref{Eqsimpl} versus voltage for sample 1. (b) RHE versus voltage for sample 1. Increasing the DC voltage allows to fulfill the Routh-Hurwitz criterion and makes expression \ref{EqRH} positive. The solid lines are a linear plot of the calculated points.}
\end{figure}

From Table \ref{tabrecap}, it appears that relation \ref{EqRj} is well verified for sample 1 and not for sample 2. For this last sample, this indicates that self-oscillation is possible even for low voltage drop along the NWR and that less power is dissipated by Joule heating. We also checked the validity of the relation \ref{Eqtps} for $\omega_0RC$ at different resistances values on the same sample. To achieve that, we lowered, step by step, the resistance of sample 1 by annealing the nanowire at increasing temperature. Before the first annealing, the pristine resistance was about 20 G$\Omega$ and self-oscillation took place at the first mode frequency, 20 kHz. Table \ref{tabrecap} shows a case at lower resistance value where it was the second mode at higher frequency and not the fundamental that self-oscillates because $\omega_0RC$ was then closer to 1. In this case  $\omega_0RC=$15 instead of 100 for the first mode. So our measurements confirm qualitatively the expected dependence of the self-oscillation threshold on the physical parameters.

Figure \ref{fig:RH} (b) shows the RHE calculated with Eq. \ref{EqRH} and the experimental parameters of table \ref{tabrecap} versus the applied DC voltage for sample 1. All this parameters are voltage independent, except $(\frac{\partial I}{\partial U})$ and  $\bar{U}$ that are calculated from experimental data as explained above for each voltage and $(\frac{\partial I}{\partial x})$ that has been measured for one voltage only and then extrapolated thanks to the Fowler-Nordheim expression. The RHE increases linearly and changes sign for a certain voltage that we call the Routh-Hurwitz threshold listed in Table \ref{tabrecap}. The voltage uncertainty presented on the tables includes only the scattering on the available data and essentially the noise on $(\frac{\partial I}{\partial U})$. This do not include the noise of $(\frac{\partial I}{\partial x})$ as it was measured for only one voltage. This threshold agrees very well with the threshold from the simplified model, confirming that close to the transition the two coupled electrical and mechanical equations can be replaced by the equivalent resonator of Eq.\ref{Eqsimpl}. The damping $\gamma/2\pi$ given by the simplified model (Figure \ref{fig:RH} (a)) is less than 10 Hz and roughly agrees with the data from the PSD of the total current. In the PSD, the duration of the signal limits the resolution to 5 Hz and the signal to noise ratio allows just to say that the width of the peak is equal or smaller than approximately 10 Hz. However, increasing the duration of the signal couldn't improve the resolution as will be explained in section \ref{phase}. The calculated thresholds are higher than the one obtained from the amplitude of self-oscillation and this apparent discrepancy would remain even with more accurate measurements. It comes from the fact that our model doesn't takes into account fluctuations and non-linearities. The next section will explain the reason of the existence of two thresholds.

\section{Study of fluctuations and non-linearities in the current}
\label{nonlin}
As explained in the introduction, self-oscillations are possible thanks to non-linearities in the dynamics to compensate the sign of the negative linear damping at high amplitude. Furthermore, the presence of hysteresis as observed in Fig. \ref{fig:A(Vdc)} (c) is a clear evidence of non-linear effects and indicates that the linear model presented in the previous section will miss some aspect of the underlying physics. In this section, we will study the non-linear dynamic of our NWR in the bistable regime and the self-oscillating state by measuring current fluctuations.

\subsection{Current fluctuations in the bistability regime}

The NWR fluctuations can be studied by analyzing the time dependence of the field emission total current for a fixed DC applied voltage. From this time dependent data the current probability density function (PDF) can be extracted. Independently, after an Hilbert transform, the amplitude A(t) and phase $\varphi(t)$ of the self-oscillator can be obtained. The phase data will be presented at the end of this section. The time average of A(t) has already been presented in Fig. \ref{fig:A(Vdc)} (c) and its fluctuations are connected to current fluctuations. So we will first focus on current fluctuations and its PDF. The PDF is a statistical tool representing an histogram of the different values taken by a random variable. Its shape provides information about fluctuations and probes the dynamical equation governing the system. Roughly speaking, in our case, electronic fluctuations cause field emission total current to wander away from the equilibrium position while other terms in the dynamical equation maintain this variable in its vicinity and damps this motion as stated by the fluctuation dissipation theorem.  The next subsection will give more theoretical details about the connection between the PDF and the dynamics of the system.

\begin{figure}
\includegraphics[width=5.5cm]{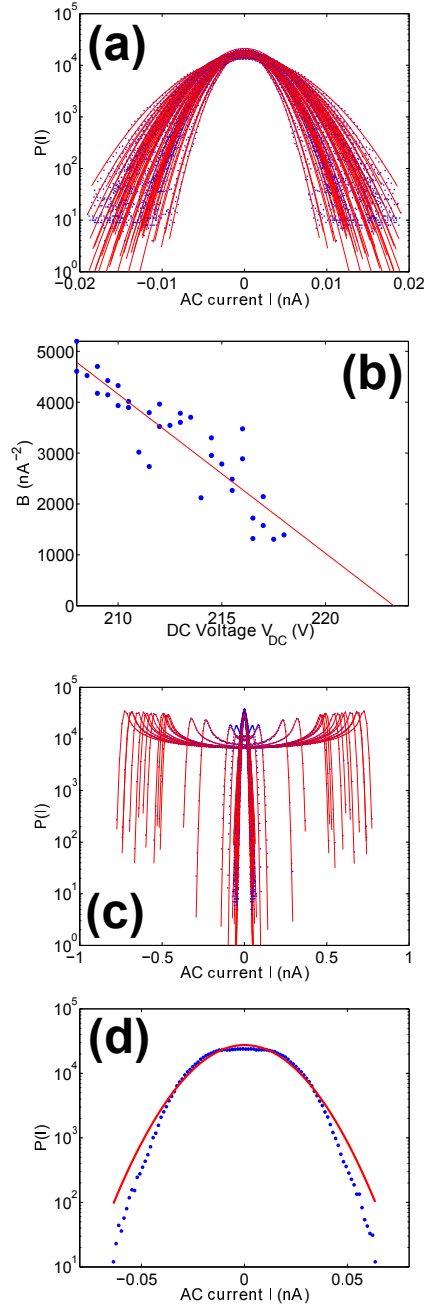}
\caption{\label{fig:pdfI} (a) Semi-logarithmic plot of the probability density function (PDF) of the AC component of the field emission current for different DC voltages for sample 2 below the self-oscillation experimental threshold. The solid lines are a fit of the data points with a Gaussian. The width increases as the experimental threshold is approached. (b) Term proportional to the linear damping of the oscillator for different DC voltage applied to sample 1. The solid line is a linear fit. (c) Semi-logarithmic plot of the PDF of the AC component of the field emission current for different DC voltages for sample 1 below and above the experimental threshold. The solid lines are a fit of the data points with a gaussian below the experimental threshold and with Eq. \ref{eq:pdfII} above the experimental threshold. (d) Semi-logarithmic plot of the PDF of the AC component of the field emission current for V$_{DC}$ = 218.5 V for sample 1. The gaussian fit in solid line is not satisfying for this voltage.}
\end{figure}

Figure \ref{fig:pdfI}(a) represents the PDF of the AC current for sample 2 (the same measurements has been performed for sample 1), filtered around the resonance peak in the PSD, for different DC voltage when the system is not self-oscillating. The PDF is Gaussian and can be fitted by the following function $exp(-BI^2)$. B decreases linearly when approaching the experimental threshold as plotted in figure \ref{fig:pdfI}(b). The data points include the measurements from the ramping up and down of the voltage. The voltage where B cancels is 223.3 V $\pm$ 0.77 V, as given by a linear fit of Figure \ref{fig:pdfI}(b). We reported this voltage as the PDF threshold in Table \ref{tabrecap}. This term will be justified in the next subsections. The voltage uncertainty is also given by the fit. This voltage is an extrapolation of the data as, once the experimental threshold is crossed (i.e. in the self-oscillating regime), the PDF is no longer Gaussian. The PDF threshold is significantly different than the experimental threshold. In the self-oscillating state, i.e. above the experimental threshold, the PDF has a totally different shape with two peaks as shown on Fig. \ref{fig:pdfI}(c).

\subsection{Existence of 2 thresholds in the bistability regime}

To interpret the shape of the PDF, the simplified linear model described by Eq. \ref{Eqsimpl} needs to be extended. We will focus on the dynamical behavior of the AC amplitude of the field emission current because it is the total current that is measured with the SED and not the position x(t). Deducing an equation for the AC current from the dynamical voltage and position equations is rather tedious due to the number of possible non linear terms. A simple phenomenological non-linear equation for the time dependent AC current defined in section \ref{lin} as $I(t) = I_{tot}(t)-\bar{I}$ is :
\begin{equation}
\label{EqnonlinI} \ddot{I}+(\frac{\omega}{Q}-f((I))\dot{I}+\omega^2I= \eta(t)
\end{equation}
where $\eta(t)$ is due to the fluctuations (thermal, shot noise ...), f(I) is responsible for the change of sign of the dissipation and the self-oscillation behavior as well as for the saturation of the AC current due to non-linear dissipative terms.

In the non self-oscillating state, the distribution of AC current due to fluctuations is obtained from the stationary solution of the corresponding Kramers equation\cite{riskenfokker}. If the zero order term in f(I) is only considered for the moment and if the fluctuations spectrum is white and constant, the distribution is Boltzmannian of the form $exp(-(m_{eff}\omega^2I^2+m_{eff}\dot{I}^2)/2k_BT_{eff})$ where $m_{eff}$ is an effective mass, $k_B$ the Boltzmann constant and $T_{eff}$ the effective temperature in our case larger than the room temperature due to electronic noise. As only the fluctuations in I and not in $\dot{I}$ are measured, this distribution can be integrated over $\dot{I}$ to get the AC current distribution. So, the distribution of I, P(I), i.e. the PDF of I, is Gaussian, to first order in the non self-oscillating state :
\begin{equation}
\label{EqP0} P(I) = P_0exp(-\frac{m_{eff}\omega^2}{2k_BT_{eff}}I^2)
\end{equation}
It can be seen, from this expression that -log(P(I)) is a measure of the potential (here parabolic) felts by the dynamical degree of freedom I.

According to the linear models of section \ref{undirect}, for a certain voltage $V^*$, the linear damping should cancel and the zero order term in f(I) should be of the form  $\omega_0/Q+\alpha(V^{DC}-V^*)$ with $\alpha>0$. Hence, in the bistable region and in the self-oscillating state higher order terms in f(I) will start to play a role. To illustrate, we used the simplest form of f(I) in the subcritical case, where a and b are supposed constant and positive :
\begin{equation}
\label{EqnonlinI2} \ddot{I}+(-\alpha(V^{DC}-V^*)+4aI^2-8bI^4)\dot{I}+\omega^2I = \eta(t)
\end{equation}
In these regimes, the amplitude A(t) of the self-oscillator and its phase, defined as before as $I(t) = A \cos(\omega t -\varphi)$ where $A>0$ and phi is a real number, are more suitable to study the dynamic of the system. So this expression is inserted into Eq. \ref{EqnonlinI2}. Then the method to solve this equation is based on a 2 time scales approach. A fast time scale of the oscillator related to $2\pi/\omega_0$ and a slower time scale related to the time evolution of the amplitude and phase. After separation of the cos and sin terms, we obtain dynamical equations reformulated in the rotating frame\cite{Rytov1956}:

\begin{eqnarray}
\label{eq:tourant}
2\omega\frac{\partial A}{\partial t}= \omega A(\alpha(V^{DC}-V^*)+aA^2-bA^4) +\eta_\bot(t) = F_{eff}(A)+\eta_\bot(t)\\
2\omega A\frac{\partial \varphi}{\partial t} = \eta_{//}(t)
\end{eqnarray}
where we defined $F_{eff}$ as the effective force applied on an equivalent overdamped particle at the position A(t) and
\begin{equation}
\eta(t) = \eta_{//}(t) cos(\omega t- \varphi(t)) - \eta_\bot(t) sin(\omega
t- \varphi(t))
\end{equation}

\begin{figure}
\includegraphics[width=8cm]{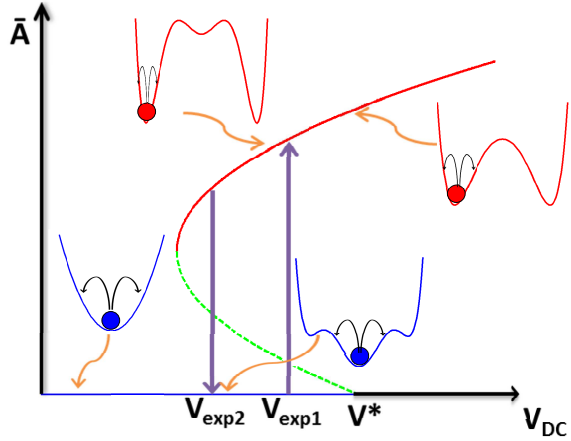}
\caption{\label{fig:dess} Diagram of the equilibrium amplitude $\bar{A}$ as function of the control parameter $V^{DC}$. The solid lines represents the stable position and the dashed line the unstable one. The effective potentials acting on the equivalent particule are represented for different regimes. The different thresholds are defined on the horizontal axes. Vertical arrows indicate the transition from one stable state to another.}
\end{figure}

A Duffing term was not included in Eq. \ref{EqnonlinI2} as it will only influence the phase and not the amplitude of self-oscillation and can be considered as a simple shift of the frequency of oscillation. The stationary solutions of this system  show that the phase can take any value and in the bistable regime, there are 3 equilibrium amplitudes : two stable $\bar{A}$ = 0 and $A_s$ given by :

\begin{equation}
A_s^2=\frac{a}{2b}+\sqrt{(\frac{a}{2b})^2+\frac{\alpha(V^{DC}-V^*)}{b}}
\end{equation}
and one unstable $A_u$:
\begin{equation}
A_u^2=\frac{a}{2b}-\sqrt{(\frac{a}{2b})^2+\frac{\alpha(V^{DC}-V^*)}{b}}
\end{equation}
with $0 < A_u < A_s$. For a voltage higher than what we will call the linear threshold $V^*$ (i.e. the voltage where the linear damping cancels), $A_u$ and $\bar{A}$ = 0 merge into one unstable position. So, in a subcritical self-oscillation transition, close to the linear threshold, several equilibrium positions coexist for a given range of the control parameter. In our experiment, the control parameter is the DC voltage. Fig. \ref{fig:dess} represents the different values of $\bar{A}$ for different values of the control parameter with the typical shape of the effective potential acting on the equivalent particule in the rotating frame.
For low voltage, this system is in a parabolic potential and it is not self-oscillating. In the multi-stability region, the system state can jump from one equilibrium position to another, for instance from a non self-oscillation state where $\bar{A}$ = 0 to a self-oscillating one, with the help of fluctuations to overcome the barrier between this states.
In particular, the non-self-oscillating state should be stable till $V^*$, as predicted by a linear model, but fluctuations let the system jump to a self-oscillating, more stable state (i.e. with a larger activation barrier) at $V_{exp1}<V^*$. So two thresholds can be defined for the system : $V_{exp1}$ where the NWR jumps into self-oscillation and $V^*$ where the NWR would jump in the absence of fluctuations. Once in the self-oscillation state, it can remain in this state till a voltage $V_{exp2} < V_{exp1}$.

Now, the PDF of A can be obtained from the corresponding Fokker Planck equation\cite{riskenfokker}.
\begin{equation}
\label{FP} \frac{\partial W}{\partial t} = D \frac{\partial^2
W}{\partial A^2}-\frac{1}{2\omega}\frac{\partial }{\partial
A}(WF_{eff})
\end{equation}
where W(A,t) is the probability for the self-oscillator to have an amplitude A at the time t, D is the diffusion coefficient related to the noise $\eta$. The PDF of A, $\bar{W}(A)$, is the stationary solution of the Fokker Planck equation :
\begin{equation}
\label{beta}
\bar{W}(A) = \bar{W}_0\exp(-a_0(V_{dc}-V^*)A^2+a_1A^4-a_2A^6)\approx \bar{W}_s \exp(-\beta(A-\bar{A})^2)
\end{equation}
where $a_0$, $a_1$ and $a_2$ are related to $\alpha$, a and b, $\bar{W}_0$ and $\bar{W}_s$ are some prefactors. W has been expanded around $\bar{A}$ the amplitude of self-oscillation. The predicted distribution of A is also Gaussian to first order. $\beta$ is a parameter that depends on the previous coefficients and is related to the inverse of the Gaussian width of the distribution. The relationship between the PDF of I and the PDF of A is given by :
\begin{eqnarray}
\label{eq:pdfII}
P(I) = \int P(I,A)\bar{W}(A)dA \\
P(I,A) = \frac{1}{\pi}\frac{1}{\sqrt{A^2-I^2}}
\end{eqnarray}%
where P(I,A) is the usual probability density of finding an oscillator at the "position" I when its motion is a cosine with an amplitude A.

\subsection{Analysis of the PDF in the bistability regime}
Experimentally, $V_{exp1}$ corresponds to the voltage where the average AC current amplitude $\bar{A}$ jumps for a voltage up-sweep in Fig. \ref{fig:A(Vdc)} (c) (reported as the experimental threshold in Table \ref{tabrecap}) as well as the voltage where the shape of P(I) changes abruptly in Fig. \ref{fig:pdfI} (c). Similarly, $V_{exp2}$ corresponds to the abrupt change for a voltage down sweep. From the previous model, the Gaussian shape of the PDF in absence of self-oscillation has been justified and B can be identified to $m_{eff}\omega^2/2k_BT_{eff}$. According to the fluctuation dissipation theorem, the fluctuations are equal to the product of the dissipation by the effective temperature. Then the effective temperature is inversely proportional to the dissipation. As the dissipation comes from the damping term in Eq.\ref{EqnonlinI}, B is proportional to the linear damping. As expected, figure \ref{fig:pdfI}(b) shows that the damping is decreasing when approaching the self-oscillation linear threshold. This measurement confirms the linear dependence of the linear damping close to self-oscillation transition predicted by the RHE and the simplified model (Fig. \ref{fig:RH}). So, although the experimental measurements based on the simplified linear model and the RHE can not predict the exact linear threshold because of fluctuations, a signature of this predicted voltage where the damping cancels is detectable in the total current fluctuations with no adjustable parameters. The PDF threshold  is an accurate experimental measurement of the linear threshold $V^*$.

Close to this threshold, a departure from a Gaussian fit starts to be visible in the PDF due to a higher order term (see Fig. \ref{fig:pdfI} (d)). Though too few data points in voltage were taken to determine whether this term is constant or not. Theoretically we could observe this non-linear term for lower voltages but it would require an acquisition time incompatible with the field emission total current stability and the 1/f noise because for lower voltage, the damping is higher, so the parabola coefficient is stronger and high amplitude fluctuations that can sense a higher order term become less probable.

It appears in this subsection that measuring the PDF gives the value of the experimental threshold $V_{exp1}$ as well as the Routh-Hurwitz (or linear) threshold $V^*$ : $V_{exp1}$ is obtained when the PDF shape changes abruptly whereas $V^*$ is deduced by extrapolating B to the voltage where it cancels (i.e. the PDF threshold). The PDF is a more powerful tool to study self-oscillations than for instance the PSD. This comes from the fact that i) For the same temporal measurement of the total current the resolution of the resonance peak in the PSD is insufficient to extract the evolution of the damping as stated in section \ref{verifexp} and ii) the PDF rely only on the AC current amplitude and so is insensitive to the phase noise contrary to the PSD.\\

\subsection{Amplitude of current in the self-oscillating regime}
\label{modnonlin}

\begin{figure}
\includegraphics[width=8cm]{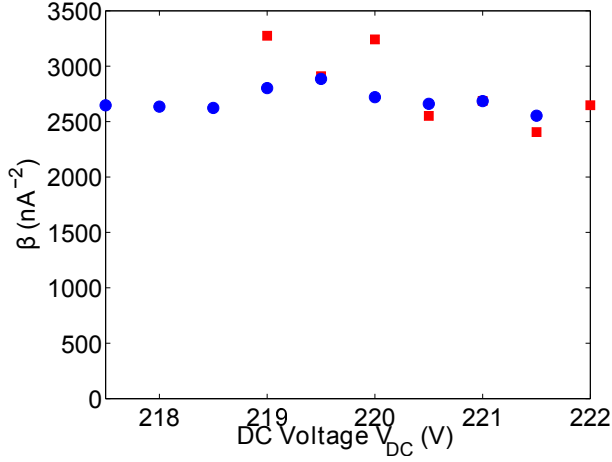}
\caption{\label{fig:nonlincoef}Dependence of $\beta$ on voltage for sample 1 (square : upward voltage sweeping, circle : downward voltage sweeping.}
\end{figure}

Equation \ref{eq:pdfII} gives the typical shape of the PDF of I with its two peaks as observed in Fig. \ref{fig:pdfI}(c) above threshold. It is remarkable that the form of the PDF is very different for a self-oscillator compared to a noise driven resonator, whereas the peak in a PSD is Lorentzian above or below the self-oscillation threshold. The shape of the PDF of I and the dependence of the AC amplitude with the voltage is roughly what is expected for such a simplified model. In contrast, the dependence of its fluctuations during self-oscillation is rather unexpected. This dependence is obtained by fitting the PDF of the AC current above threshold with equations \ref{beta} and \ref{eq:pdfII} and extracting $\beta$ as plotted in figure \ref{fig:nonlincoef}. It appears that $\beta$ is insensitive to the voltage while the existence of the $a_0(V_{dc}-V^*)$ term in $\bar{W}(A)$ should induce a dependence. We conclude that the non-linear behavior of the self-oscillator controlling the saturation as well as the span of the hysteresis region cannot be described by our simplified first order non-linear model probably because of the high amplitude of the AC current and vibration or because the terms a and b have a voltage dependence.

\subsection{Phase of the self-oscillator}
\label{phase}
The phase is the parameter that determines the stability of a self-oscillator for its use as a time base. The Allan deviation is used to quantify this stability as it quantifies the stability on different time scales. We computed this Allan deviation from the argument of the Hilbert transform (i.e. the phase $\varphi$ of the AC current) :
\begin{equation}
\label{Allan dev}
\sigma(\tau) = \sqrt{\frac{1}{2}\sum_i\frac{1}{N-1}(<\dot{\varphi}(t_i)>_\tau-<\dot{\varphi}(t_{i+1})>_\tau)^2>}/<\dot{\varphi}>
\end{equation}
where the $t_i$ are different times separated by a time $\tau$ and the notation $<>_\tau$ means a time average during a time $\tau$ around $t_i$.
\begin{figure}
\includegraphics[width=8cm]{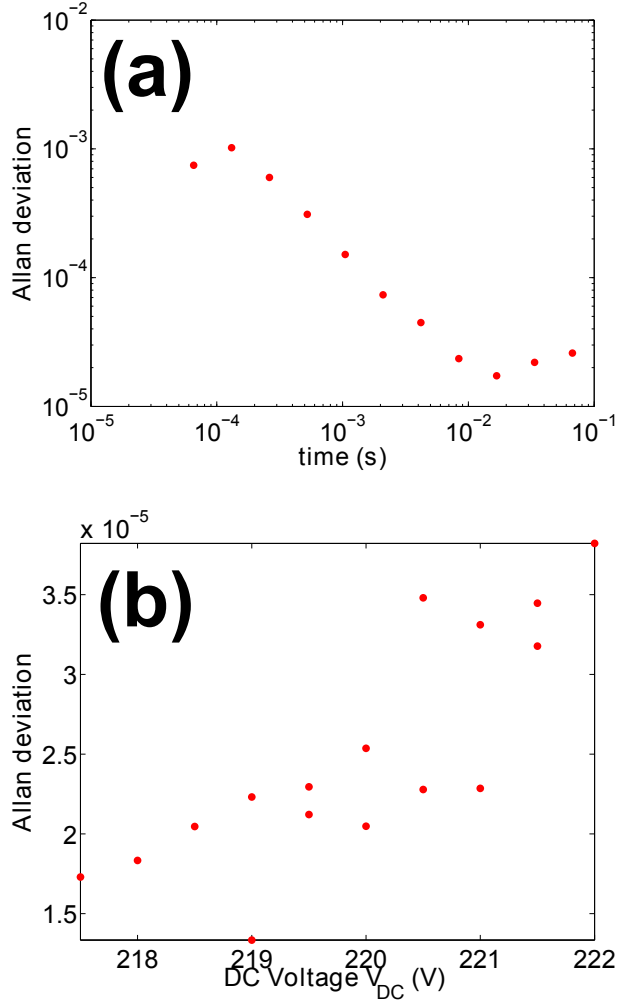}
\caption{\label{fig:allandev} (a) Allan deviation for sample 1 for a DC voltage of 217.5 V in the bistability region. (b) Minimum of the Allan deviation for sample 1 versus DC voltage.}
\end{figure}
The minimum of the Allan deviation is generally above 10 ms for both samples. Fig. \ref{fig:allandev}(a) shows a typical Allan plot for sample 1. Due to the instability of the field emission process and the 1/f noise observed in the PSD of the emission total current, the Allan deviation increases for times above several tens of ms. This long term phase drift will make the width of the resonance peak in the PSD larger for longer duration of the signal although the frequency resolution of the Fourier transform will increase. That's why, the determination of the intrinsic damping of the resonator from the PSD is limited even for long recording time. Fig. \ref{fig:allandev}(b) plots the minimum of the Allan deviation of sample 1 for each DC voltage. Our best Allan deviation is 1.3$\cdot$10$^{-5}$ and the smallest value appears close to the self-oscillation threshold (i.e. for the lowest amplitude of oscillation). This value is roughly 10 to 50 times worse than what is observed in self-oscillating NEMS with external feed-back (for instance the allan deviation in Ref.~\onlinecite{chaste2012} is 2$\cdot$10$^{-6}$ for carbon nanotubes). However our measurements are performed at room temperature while in the literature the Allan deviation in NEMS is usually given at cryogenic temperature. Our samples might well reach the state of the art of NEMS if the measurements were made at a lower temperature.

\section{Conclusion}
In this article, we performed an experimental and theoretical study of a self-oscillating field emission NEMS with intrinsic feedback and measured numerous physical parameters controlling the phenomenon. A simple linear model was shown to predict qualitatively the cancelation of the damping close to the self-oscillation threshold. We demonstrated that the amplitude of self-oscillation is quite large and comparable to the DC signal flowing through the circuit. Although hysteresis in the IV characteristics can be removed, the system remains intrinsically subcritical with abrupt jumps in the self-oscillation amplitude. The PDF of the AC current has been used to demonstrate the coexistence of 2 thresholds in the system. One related to the cancelation of the linear damping and a lower one depending on the noise amplitude. The PDF is more useful than the PSD to study self-oscillation thresholds. The stability of the oscillator is reasonable for a NEMS but remains too low for practical purposes. Due to the high amplitude of vibration, the non-linear dynamics of the system cannot be described by a simple model and would require a deeper theoretical analysis. This work opens the door for the study of the synchronization of such highly non linear self-oscillators.

\begin{acknowledgments}
This work was supported by French National Research
Agency (ANR) through its Nanoscience and Nanotechnology
Program (NEXTNEMS, ANR-07-NANO-008-01) and Jeunes
Chercheuses et Jeunes Chercheurs Program (AUTONOME,
ANR-07-JCJC- 0145-01). The authors acknowledge the
"Plateforme Nanofils et Nanotubes Lyonnaise" of the university
Lyon1.
\end{acknowledgments}

\end{document}